\documentclass[prl,twocolumn, 11pt, reprint,nofootinbib]{revtex4-1}
\usepackage{graphicx}\usepackage{color}
\def\simpropto{\lower.2ex\hbox{$\; \buildrel \propto \over \sim \;$}}
\def\simlt{\lower.5ex\hbox{$\; \buildrel < \over \sim \;$}}
\def\simgt{\lower.5ex\hbox{$\; \buildrel > \over \sim \;$}}
\def\bea{\begin{eqnarray}}
\def\eea{\end{eqnarray}}
\begin{document}

\title{ Pulsar scintillation patterns and strangelets}

\author{  M. \'Angeles P\'erez-Garc\'ia$^1$~\footnote{mperezga@usal.es}, Joseph Silk$^2$~\footnote{silk@iap.fr} and Ue-Li Pen$^3$~\footnote{pen@cita.utoronto.ca}
}

\affiliation{ $^1$ Department of Fundamental Physics and IUFFyM, \\University of Salamanca, 
Plaza de la Merced s/n 37008 Salamanca,\\
$^2$ 
Institut d'Astrophysique,  UMR 7095 CNRS, Universit\'e Pierre et Marie Curie, 98bis Blvd Arago, 75014 Paris, France\\
Department of Physics and Astronomy, The Johns Hopkins University,
Homewood Campus, Baltimore MD 21218, USA\\
Beecroft Institute of Particle Astrophysics and Cosmology, Department of Physics, University of Oxford, Oxford OX1 3RH, UK\\
$^3$ Canadian Institute for Theoretical Astrophysics, University of Toronto, 0N M5S 3H8, Canada}

\date{\today}

\begin{abstract}
We propose that interstellar extreme scattering events, usually observed as  pulsar scintillations, may be caused by a coherent agent rather than the usually assumed turbulence of $\rm H_2$ clouds. We find that the penetration of a  flux of ionizing, positively charged strangelets or quark nuggets into a dense interstellar hydrogen cloud may produce ionization trails. Depending on the specific nature and energy of the incoming droplets, diffusive propagation or even capture in the cloud are possible. As a result, enhanced electron densities may form and constitute a lens-like scattering screen for radio pulsars and possibly for quasars.
\end{abstract}

\maketitle

A variety of scintillation phenomena observed from pulsars and quasars require interstellar scattering screens that contain compact regions of high electron density. 
These  include quasar Extreme Scattering Events (ESE)  \cite{fiedler87, fiedler94}, pulsar parabolic arcs \cite {cordes04, walker04} and Galactic Center scattering of OH maser sources \cite{lazio98, goldreich06}.
Many of these phenomena, in particular the ESEs, require enhanced electron density regions of A.U. size ($\sim1.5\times 10^{13}$ cm). The overpressure $P/k_B\approx 10^6-10^8 \,\rm K\,cm^{-3}$, as estimated from typical temperatures $T\sim 10^4\, \rm K$ and particle densities $n\sim 10^2-10^4\, \rm cm^{-3}$, is difficult to explain in any conventional scattering screens embedded in dense molecular $\rm H_2$ clouds. The only plausible environment where such pressures might be attained would be in the dense cloud cores (cf.  the compact ionized cloud model developed by Walker {\cite {walker01, walker07}). The required source properties in this latter model require a significant mass in extremely dense cold gas clumps to source the ionized gas. The stability of such cold clumps is questionable, although exotic models have been proposed \cite{revaz09}.

There is some direct evidence that the ionized clouds are highly elongated \cite{walker09}. It is an already well-known fact that plasma lenses result from electron (over-) under-densities. Electron over-densities result in a faster phase velocity, corresponding to a concave (divergent) optical lens, while under-densities are associated with a convergent lens \cite{pen}. Quantitative lens models for the ESEs \cite {walker01} estimate electron column densities $N_e \sim\,10^{15}\,\rm cm^{-2}$. 

Here we are interested in estimating the effect of the enhancement of the electron column density on A.U. scales as a result of the formation of ionized trails in  $\rm H_2$ (molecular) and HI  (atomic) hydrogen clouds caused by an external agent. We propose a mechanism that provides an alternative to postulating the existence of  controversial clumps of dense molecular hydrogen, and naturally generates pervasive ionized trails in dense interstellar cloud cores. Our model invokes strangelets \cite{strangelets} (also known as {\it nuclearites}), finite droplets of quark matter with non-zero strangeness fraction and slightly charged. They are currently being searched on earth \cite{lu} as final products in heavy ion collisions, with the ALICE experiment at the LHC, in the CDMSII, under the form of light ionizing particles, or in the AMS-02 mission. It is expected that these quark droplets can be naturally generated by a series of different astrophysical events where  a nucleon-quark deconfinement transition may take place, e.g. neutron star (NS)  collisions, NS or black hole combined binary mergers or in  NS to quark star (QS) conversions. This latter process may be induced by internal heating due to dark matter annihilations \cite{perez10} (under the assumption of a Majorana particle candidate) and even leave observable traces in the pulsar distribution \cite{perez12} or in the emission of very short gamma ray bursts (GRBs) with typical time scales $T_{90}\lesssim 0.1$ s \cite{perez13} detectable with modern projected missions \cite{gamma}. Due to the large gravitational and nuclear binding energies released in the transitioning process, a mass ejection episode is expected, possibly seeding the interstellar medium with a fraction of strangeness-carrying lumps of matter formed during the phase of nucleon deconfinement. Energetics show that the measured short GRBs isotropic equivalent photon emission value $E_{\gamma\, iso}\sim 10^{48}-10^{52}$ erg is compatible with  relativistic mass ejecta $M_{\rm ej}\lesssim10^{-4}M_{\odot}$  able to consistently produce observable gamma rays. It has been actually proposed that quark matter droplets might partially populate cosmic ray (CR) primaries (see e.g. \cite{madsen06, shaham12, kotera}). 

In this context, let us consider a cloud of mixtured $\rm H_2$ and HI. Typical ionization reactions are of the form, $\rm X+H_2 \rightarrow X+H^+_2+ e^-$, or $\rm X+ HI  \rightarrow X+ H^+ +e^-$, where $\rm X$ is the incoming charged strangelet. In addition, electron capture by the positively charged strangelet could be, in principle, possible \cite{berger}. The energy needed to ionize a hydrogen atom (molecule), initially in the ground state is  $\rm I(HI)=13.6$ eV ($\rm I(H_2)=15.6$ eV). In astrophysical CGS units a more practical conversion factor $1\,\rm eV=1.62\, \times10^{-12}$ erg is used.
The dimensions of the molecular cloud (MC) vary but the denser regions with $n\sim10^{4\div5}$ $\rm cm^{-3}$ are typically less than $\sim$ 1 pc. Assuming for the cloud core $\rm R_C\sim0.1\, \rm pc\sim 3\,\times10^{17}$ cm, then the  core volume is $\rm V_C\approx 4 \pi R^3_C/3 \simeq 1. 1\, \times10^{53}$ $\rm cm^3$. Taking in the cloud core $n_0\sim10^{4}$ $\rm cm^{-3}$, the number of particle species (HI, protons, electrons) is  accordingly given by $n_0\,\rm V_C\simeq 1.1\,\times 10^{57}$. In addition, non-vanishing magnetic fields are expected in the MC and can be parametrized \cite{crutcher} as $B\sim 100 (\frac {n}{10^4\,\rm cm^{-3}})^{1/2}$ $\mu$G.
In general, other ionizing agents, such as CRs can cause ionization of $\rm H_2$ or HI with a rate  $\xi^{HI}\sim 10^{-15}\, \rm s^{-1}$ \cite{spitzer} or $\xi^{H_2}\sim 10^{-17}\, \rm s^{-1}$ \cite{glassgold}. A more exotic flux of potentially ionizing electrically charged strangelets through the cloud  will depend, on the one hand, on the time and spatial distribution of their astrophysical sources (besides a possible primordial background) and, on the other hand, on their peculiar nature.

We will consider a simplified model with emission of lumps of baryonic number $\rm A$ and mass $m_{A}\lesssim Am_{\rm N}$, where $m_{\rm N}$ is the nucleon mass. The detailed mass formula can be obtained from existing calculations \cite{berger}. The strangelet number production rate in the astrophysical ith-process  will be given by $dN_{{\rm A}, i}/dt=\eta_i \dot M_{{\rm ej}, i}/m_{A}$. ${\dot M_{{\rm ej},i}}$ is the mass rate and $\eta_i$ is the efficiency of strangelet ejection for the ith-process involved, respectively. For example, for NS collisions it is expected that $M_{\rm ej}\sim (10^{-5}-10^{-2})  M_{\odot}$ while in a NS merger the mass ejection is $M_{\rm ej}\sim (5\,\times10^{-4}-7\,\times10^{-3}) M_{\odot}$ for equal-mass binaries with total mass $m = 2.7M_{\odot}$ \cite{Hotokezaka}. In a NS to QS transition capable of emitting a GRB, it is expected that $M_{\rm ej}\lesssim10^{-4} M_{\odot}$.  According to \cite{perez13} the rate of these transitions is $R_{\rm NS-QS}\simeq (8\,\times 10^{-4}-3\,\times10^{-3}) R_{\rm SNtII}$ being $R_{\rm SNtII}\simeq 10^{-2}$ $\rm yr^{-1}$ the rate of type II supernovae in our galaxy. 

Having enumerated the possible processes that may constitute a source of the strangelet flux, for the rest of this work for practical purposes we will consider a {\it generic source} where the deconfinement transition can take place with galactic appearance rate $R={\bar R}\simeq 10^{-5}$ $\rm yr^{-1}$ and ejected mass $M_{\rm ej}={\bar M_{\rm ej}}\simeq 10^{-5} M_{\odot}$. The ejected mass rate is then ${\dot M_{\rm ej}}=R\,M_{\rm ej}\simeq 10^{-10} M_{\odot}\, \rm yr^{-1}$. Since the efficiency of strangelet production depends on the so far unknown details of the engine model, we will consider that only a small fraction $\eta \sim 10^{-2}$ is ejected under exotic form \cite{perez13}. Nevertheless,  strangelets should be emitted with A-values larger than a critical stability value \cite{amin}, $A>A_{\rm min} \simeq 10^1-10^2$ so that they can possibly decay to the lightest energetically stable $A_{\rm min}$-species. 

The peculiar nature of the quark droplets is highly uncertain but it is usually assumed that their charge is small and distributed positively on the surface \cite{madsen06}. For ordinary strangelets $Z\simeq A^{1/3}$ \cite{madsen_cfl} while for CFL (color-flavour-locked) strangelets $Z\simeq 0.3 A^{2/3}$. Even smaller charge-to-mass ratios are energetically possible for intermediate masses, $A\sim 10^2-10^{18}$, assuming a strong coupling $\alpha_S=0.9$ \cite{madsen}. For example, for $A\simeq 10^9$, $Z/A\simeq 10^{-4}$ while larger strangelets $A\simeq 10^{18}$ have $Z/A\simeq 10^{-7}-10^{-3}$ and even $Z/A<0$. In this work we constrain strangelets  to have $Z\geq1$.

To estimate the strangelet production number rate we are going to consider scintillation from a galactic emitting generic source located at a distance $d_{\rm SO}\lesssim 10$ kpc from an observer. As an example, pulsar simulations yield a spatial distribution peaking about galactocentric radii $\sim5$ kpc and vanishing beyond $\sim$12 kpc \cite{Hart}. We will also discuss the effect from a possibly nearby pulsating source \cite{brisken} \cite{bhat} at $d_{\rm SO} \sim 600$ pc. Then, the A-sized strangelet number galactic prodution rate at the generic source is given by
\begin{equation}
\frac{dN_{A}}{dt}=2\, \times10^{45} \left(\frac{\eta}{0.01}\right)
\left(\frac{{\dot M_{\rm ej}}}{10^{-10} M_{\odot}\, \rm yr^{-1}}\right) \frac{f_{S}(Z, \beta) }{A} \, \rm yr^{-1}.
\label{eq1}
\end{equation}
It is expected that a possible emission distribution function at the source $f_{S}(Z, \beta)$ can modulate this rate.  We will not consider this refinement here, and in what follows we will assume $f_{S}(Z, \beta)\sim 1$.

 Due to the fact that strangelets are electrically charged they will diffuse in the magnetized medium and the effective distance traveled over the rectilinear distance $d_{\rm s}$, is obtained as $l(d_{\rm s})=d_{\rm s}^2c/(2D)$ in a corresponding diffusive time $t_{\rm diff}\sim l(d_{\rm s})/c$. For the  diffusion coefficient, $D$, in the galactic halo we take \cite{Blasi12_1}} $D(E) = 1.33\times10^{28}H_{\rm kpc}[E/(3Z\,{\rm GeV})]^{1/3}$\,cm$^2\,$s$^{-1}$, where $H_{\rm kpc}\equiv H/(1\,{\rm kpc})$ is its height. In the MC larger values of the magnetic field are assumed and following \cite{Gabici}, we take $D_{\rm MC}(E) \simeq 1.7 \times 10^{27}[E/(Z\,{\rm GeV})]^{1/2}[B/10\,{\rm \mu G}]^{-1/2}$ cm$^2\,$s$^{-1}$ with an averaged value over the MC of $B\sim10 \,\mu$G.

Typically, the strangelet ejection energy at a transitioning source allows Lorentz factors bounded by a saturation value, $\gamma \lesssim \gamma_{\rm sat}$ with $\gamma_{\rm sat}\sim20-1000$ \cite{perez13}. Correspondingly, the kinetic energy is $T\sim (\gamma-1)A\rm \, GeV/c^2$ so that we will assume $T\sim AT_0 \sim A {\rm TeV}$ droplets with a moderate $A\gtrsim A_{\rm min}$  and $Z$ charge. In such scenario and in the universe lifetime, $\tau_{\rm u}$, $N_{\tau_{\rm u}}\sim R \tau_{\rm u}\sim 10^{5}$ sources  could be expected at $d_{\rm SO}\lesssim 10$ kpc. In that case  the {\it unscreened} diffusive flux is $F\simeq F_0 {N_{\tau_u}}$ and
\begin{equation}
\small
F_0 \simeq \frac{dN_{A}}{dt} \frac{1}{ 4 \pi l(d_{\rm SO})^2} \sim 2.3\, \times10^{-16} Z^{-2/3} A^{-1/3} \, \rm cm^{-2} sr^{-1} s^{-1}.
\label{eqf}
\end{equation}
We must note, however, that some sources may be closer than the assumed $\sim 10$ kpc. In the case that $d_{\rm SO}\sim 600$ pc \cite{brisken} then there are additional volume $\frac{V_{600\rm\, pc}}{V_{10\rm\, kpc}}\sim 2.1\times 10^{-4}$ and distance $\left(\frac{l(10\,\rm kpc)}{l(600\,\rm pc)}\right)^2 \sim 7.7\times 10^{4}$ factors yielding an estimate one order of magnitude larger than previous value (we have roughly assumed averaged source distribution in the halo).

As a comparison, for strangelets being currently searched in neutrino telescopes on earth, there is a lower limit $A\gtrsim 10^{13}$, our estimates yield in that case $F\sim 1.1 \, \times10^{-15} \, \rm cm^{-2} sr^{-1} s^{-1}$. We will restrict to  $Z=1$, $A\sim 10^3$ droplets  but larger $A$ are allowed if they are less energetic  remaining below observational CR bounds $\sim 10^{20}$ eV. If the source is nearby by chance there are presently competitive limits from experiments such as ANTARES \cite{antares}, who report  a testing capability flux $F_{\rm ANTARES} \sim 2\, \times10^{-14} \,\rm cm^{-2} sr^{-1} s^{-1}$ for $A\gtrsim 10^{13}$, or IceCube-22, who report  $F_{\rm IC22} \sim 10^{-18} \,\rm cm^{-2} sr^{-1} s^{-1} $  for $A\gtrsim10^{17}$ nuclearites \cite{ice}. It is however uncertain whether such large-A droplets can arrive on earth without suffering spallation or decay processes \cite{kotera}. 

Once the nuclearites are produced it may happen that their diffusive trajectories intersect with a MC. These are accumulated in the midplane of the galactic disk although we suppose that may also be present at higher latitudes. If the Larmor radius is comparable to the typical coherence length of the galactic magnetic field, $r_L\lesssim l_{\rm c}$, $l_{\rm c}=10-100\,\rm pc$ \cite{Han08}, then a strangelet will suffer an accumulated deflection that can be estimated from random walk approximation as,
\begin{equation}
\small
\theta (T)\simeq 5.4^\circ \left(\frac{l_{\rm c}}{100\,\rm pc}\frac{ d_{\rm SC}}{10\,\rm kpc}\right)^{1/2} \left(\frac{Z/A}{10^{-7}}\right) \left(\frac{B}{ 1\,\mu G}\right) \left(\frac{1\,TeV}{T }\right) .
\end{equation}
Thus, propagation effects in the galaxy or MC for moderate $Z/A$ will not make  possible to determine the line-of-sight direction to a distant emitting source.

In this scenario one can estimate $N^{C}_A$,  the effective number of ionizing $A-$sized strangelets traversing the MC core in its lifetime, $\tau_C\sim 10^7$ yr, as 
$N^{C}_A\simeq \eta_{\Omega} \pi R^2_C \int_0^{\tau_C} F dt$. Since beaming is observed in some very short GRBs (expected in this scenario \cite{perez13}) with opening angle $\theta_\mathrm{j}\sim 1^{\circ}-30^{\circ}$, an efficiency factor $\eta_{\Omega}\simeq \frac{4 \pi} {\Omega} \sim 1-\rm cos\, \theta_\mathrm{j}$ has been introduced. Due to the deflection effects we expect  typically $\eta_{\Omega}\simeq 0.1-1$. Using values $d_{\rm SC}\sim d_{\rm SO}\sim10$ kpc then 
\small
\begin{equation}
N^{C}_A \simeq 2\times 10^{38} A^{-1/3} Z^{-2/3} \left(\frac{\eta_{\Omega}}{0.1}\right) 
\left(\frac{R_C}{0.1\,\rm pc}\frac{10\,\rm kpc}{d_{\rm SC}}\right)^2\left(\frac{\tau_{C}}{10^7\, \rm yr}\right).
\end{equation}
\normalsize

If closer sources are considered \cite{brisken} then the MC would be illuminated,  effectively, by an order of magnitude larger amount of particles as discussed previously and even higher if the source was inside the MC \cite{kotera}. 

As the slightly charged and heavy strangelet traverses the MC with a diffusive behaviour, continuously loses kinetic energy. Strangelets of net effective charge $Z_1$ may alter the cloud  ionization fraction either by ionizing hydrogen or capturing electrons so that they change its own incident state of charge, as measured for massive ions on gas \cite{liu}. For low velocities $(\beta=v/c)$ this may be important as it diminishes the ionizing power from the bare charge $Z$ to an effective $Z_1=Z(1-e^{-(0.95 \beta/\alpha Z^{2/3}})$ \cite{pierce}. 

\begin{figure}[tbp]
\begin{center} 
\includegraphics [angle=-90,scale=0.5]{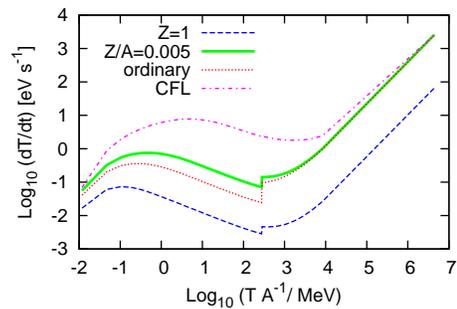}
\caption{Logarithm of the kinetic energy loss rate in the cloud as a function of the logarithm of the incoming ($A=10^3$) strangelet kinetic energy per baryon number. Several charge states are shown and a CR proton case with $Z=1$ is depicted for comparison.}
\label{Fig2}
\end{center}
\end{figure}

Interaction in the cloud may arise due to a variety of processes \cite{madsen}\cite{berger}. The strangelet (kinetic) energy loss per unit length due to ionization and pion production can be obtained from the stopping power $\frac{dT}{dx}=\frac{dT}{dt}\frac{1}{\beta c}$. At lower energies, ionization is the most important process, and for an incoming particle with effective charge $Z_{1}$ and velocity $\beta$ the energy loss rate reads \cite{book}
\begin{equation}
\frac{dT}{dt}=-1.82 \,\times10^{-3}\left(\frac{n}{10^4 \rm\, cm^{-3}}\right) [Z^2_1 \Psi(Z, \beta)+ \Xi(A, \beta)]\, \,\rm eV/s,
\end{equation}
where
\begin{equation}
\Psi(Z, \beta)=\frac{[1+0.0185\, \rm ln(\beta) \theta(\beta-\beta_0)]} {\beta_0^{3}+2\beta^3}{2\beta^2},
\end{equation}
and
\begin{equation}
\Xi(A, \beta)=0.72A^{0.53}\gamma^{1.28}\theta(\gamma-1.3),
\end{equation}
$\theta (x)$ is the Heaviside function, $\gamma^{-1}=\sqrt{1-\beta^2}$ and $\beta_0=0.01$ is the electron orbital velocity. This function is shown in Fig. \ref{Fig2} versus the strangelet kinetic energy per baryon number for an $A=10^3$ case. Different amounts of charge $Z/A=5 \,10^{-3}$ (solid), ordinary (dotted) and CFL (dot-dashed line) strangelets and a CR proton case (dashed line) have been considered. For relativistic strangelets most of the energy is kinetic and they show an enhanced ionizing power with respect to the linear case and a change in slope around $T/A\sim500$ MeV, due to pion production, for larger energies.

To estimate the hydrogen ionization rate by strangelets, we consider the number density of old NSs in our galaxy assuming $N_8 \sim 10^8$ NSs in a halo radius $\sim10$ kpc. This yields $n_{\rm oldNS}\sim \frac{10^8}{(10\,\rm kpc)^3}\sim 10^{-4}\rm pc^{-3}$. Note that MC density is $n_{\rm MC} \sim 10^{-6}\,\rm pc^{-3}$. In the lifetime of the MC from a rate of conversions $R\sim 10^{-5} \rm yr^{-1}$  the conversion fraction is $f_{\rm conv}=\tau_{ C} R/N_8\sim10^{-6}$. Therefore conversions are rare non-repeating events as naturally arises in this scenario.

Having estimated the incoming strangelet  flux at distance $d_{\rm SC}$ and the effective number of strangelets traversing the core, $N^C_A$,  the injection rate, $S$, must take into account the fewer conversions in the MC lifetime $S \sim F \frac{N_{\tau_C}}{N_{\tau_u}}$ where $N_{\tau_C}=R \tau_C$. 

\begin{figure}[tbp]
\begin{center} 
\includegraphics [angle=-90,scale=0.5]{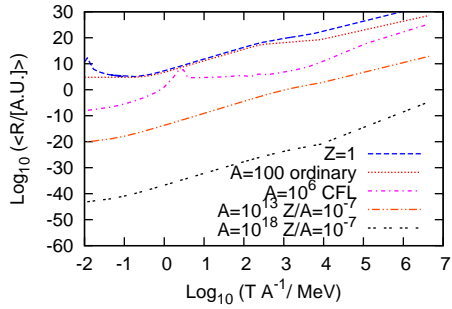}
\caption{Logarithm of the strangelet range in the cloud as a function of the logarithm of the incoming kinetic energy per baryon number. See text for details. }
\label{Fig3}
\end{center}
\end{figure}

As the rate depends on the nature and energetics of the droplet it thus influences the strangelet density in the MC. For example, taking a diffussion escape time for a $\sim$few ${\rm TeV}/A$ strangelet entering the MC, $t_{\rm diff}\sim l(R_C)/c\sim0.027\,(Z/A)^{1/2}$ yr. In that case the strangelet number density in the MC is $n_{A}\simeq S t_{\rm diff} R^{-1}_C  \sim 6.5\,\times10^{-26} Z^{-1/6} A^{-5/6}\rm cm^{-3}$. As we have seen in Fig. \ref{Fig3} some lumps may be captured in the MC lifetime and in that case $n_A$ would be a factor $\tau_C/t_{\rm diff}$ larger. For moderate A lumps, these values are much lower  than the CR density at the  low-energy break in the spectrum of Galactic CRs at $\sim1$ GeV, computed from the CR flux $F_{\rm CR} \sim 10^4 \,\rm m^{-2}\,s^{-1}\,sr^{-1}$ or $n_{\rm CR}\sim 3\,10^{-11}\,\rm cm^{-3}$ \cite{nero} and at $\sim1$ TeV, $F_{\rm CR} \sim 10^{-1} \,\rm m^{-2}\,s^{-1}\,sr^{-1}$ , $n_{\rm CR}\sim 3\,10^{-16}\,\rm cm^{-3}$.

The hydrogen ionization rate due to strangelets $\zeta_A^H$ {\it averaged over the MC}, is now estimated to be $\zeta_A^H =|dT/dt|\,n_{\rm A} \langle I \rangle^{-1} n^{-1}$, per H atom. Assuming $\langle I \rangle\simeq36$ eV per H ionization and heavy strangelets with $\gamma\sim 10^3$ then $\zeta_A^H =1.2\times 10^{-25} [Z^{1.83} A^{-5/6} +5\times10^{3}  A^{-1/3} Z^{-1/6}]\,s^{-1}$.
The strangelet mean free path or average range can be calculated as $R \sim\int dT\left(\frac{dT}{dx}\right)^{-1}$. If strangelets have charge $Z>1$ their range will be smaller since they provide a more efficient ionization. In Fig. \ref{Fig3} we show the logarithm of the strangelet range in the cloud (in A. U.) as a function of the logarithm of the incoming kinetic energy per baryon number. We consider a proton CR case $Z=1$, ordinary and CFL strangelets with $A=10^2$ and $A=10^6$ respectively, and larger strangelets with $A=10^{13}$ and $A=10^{18}$ with $Z/A=10^{-7}$. As seen the cloud can effectively stop the very heavy nuggets while this is not possible for the smaller droplets even at lower kinetic energies.

Once the cloud core is populated with trails of HII (ionized hydrogen), this region  contains $\rm H^+$ ions that tend to recombine radiatively as $\rm H^+ + e^- \rightarrow H + h\nu$. The balance between ionization and recombination determines the ionized fraction. To estimate the possibility of a net ionization in the HI cloud one must compare the  ionization and recombination times, $\tau_{\rm ioniz}$ and $\tau_{\rm rec}$ respectively, so that $\tau_{\rm ioniz}\lesssim \tau_{\rm rec}$. If recombination is slower, there would be a net amount of ionized HI in the cloud to cause an electron density enhancement. In addition, there might be a fraction of net charge due to the accumulation of positively charged lumps stopped by the cloud. The recombination time, $\tau_{\rm rec}$, is calculated in the  {\it on the spot} approximation as $\tau_{\rm rec}=1/n_e\alpha^{(2)}= 3.85\times \,10^{12}n_e^{-1}T_4^{0.8}\,\, s$ where we take $T_4=T/10^4$ K and $n_e$ is the electron number density. We estimate the  average ionization time from the average number of ionizations in the cloud lifetime as $\tau_{\rm ioniz}\simeq \tau_C/N^{C}_A$. Since $\tau_{\rm ioniz}<<\tau_{\rm rec}$ then it is indeed possible to obtain net ionization.
The ionization of interest for the ESEs occurs along the trajectory, that spans the MC, but only over a trajectory width given by the product of hydrogen recombination time and sound speed, $ c_s\sim 13 T^{1/2}_4\,\rm km \,s^{-1}$,  namely at $T=T_4$ $h\sim c_s\alpha^{-1} n^{-1} \sim 33.3/n_4\,\rm A.U.$, where $n_4=\frac{n}{10^4\,\rm cm^{-3}}$. Therefore the local ionization rate is increased by a factor $\sim (R_{\rm C}/h)^2\sim 3.6 \times 10^{5}\,n_4^{2}$ at TeV energies.  The condition  $t_{rec}^{-1} =\alpha n_e=\zeta_A^H$  yields an electron density of $n_e\sim 8.4\,\times 10^{-4}\,  A^{-1/3} Z^{-1/6} n_4\, \rm cm^{-3}$. Typical ESE events have been argued to be explained by localized and dense plasma structures \cite{ne} of length comparable to trail widths of $h\sim 1.7\,\times (10^{-4}-10^{-3})R_{C}\sim5 \times (10^{13}-10^{14})$ cm for densities $n_4=10-1$ typical in dense cores. 
The enhancement in the column density coming from the ionization produced by the incoming particles is  estimated as $\Delta N_e =\int n_e(s) ds$. Then for a typical heavy species with energies of TeV that is stopped in the cloud  the accumulated enhancement over the cloud lifetime for $n_4=1-10$  is $\Delta N_e\simeq n_e R_C\sim 8.4\times10^{-4} A^{-1/3} Z^{-1/6} n_4 \, 3 \times 10^{17}\,\rm cm\sim 2.5\times 10^{14\div 15}A^{-1/3} Z^{-1/6}\,\rm cm^{-2}$ as required by these scattering events.

We have shown that pulsar scintillations may be caused by an ionization agent constituted by positively charged lumps of strange matter. Typical obtained over-densities in electron column densities  are compatible with those of quantitative ESE models. Compact A. U. sized regions are involved in the geometry of these events. Since these quark nuggets remain to be experimentally discovered further work is needed to experimentally confirm this theoretical scenario.

MAPG would like to thank the kind hospitality of Institute d'Astrophysique in Paris where part of this work was developed and the Spanish MICINN MULTIDARK, FIS2012-30926 and FIS2011-14759-E projects.  The research of JS has been supported at IAP by  the ERC project  267117 (DARK) hosted by Universit\'e Pierre et Marie Curie - Paris 6 and at JHU by NSF grant OIA-1124403.

\section{bibliography}

\end{document}